\documentclass[]{article}
\usepackage{graphics}

\begin{document}
\begin{center}
{\bf IS AMINO-ACID HOMOCHIRALITY DUE TO\\ ASYMMETRIC PHOTOLYSIS
IN SPACE?}
\bigskip\\
C. CERF\\
{\footnotesize\it PhD in Biochemistry}\\
{\footnotesize\it D\'epartement de Math\'ematique, CP 216}\\
\medskip
{\normalsize and A. JORISSEN\footnote{Author to whom all
correspondence should be addressed}}\\
{\footnotesize\it PhD in Astrophysics}\\
{\footnotesize\it Institut d'Astronomie et d'Astrophysique, CP
226}\\   
\medskip
{\footnotesize\it Universit\'e Libre de Bruxelles,} 
{\footnotesize\it Boulevard du Triomphe,}
{\footnotesize\it B-1050 Bruxelles, Belgium}\\ 
{\footnotesize\it E-mail: ajorisse@astro.ulb.ac.be}
\end{center}

\bigskip
\bigskip
\bigskip

{\footnotesize
\noindent {\bf Abstract.} It is well known that the amino acids
occurring in proteins (natural amino acids) are, with rare exceptions,
exclusively of the L-configuration.  Among the many scenarios put
forward to explain the origin of this chiral
homogeneity (i.e., {\it homochirality}), one
involves the asymmetric photolysis of amino acids present in space,
triggered by circularly polarized UV radiation. The recent observation
of circularly polarized light  (CPL) in the Orion OMC-1 star-forming
region has been presented as providing a strong, or even definitive,
validation of this scenario. The present paper reviews the
situation and shows that it is far more complicated than usually
apprehended in the astronomical literature. It is stressed for
example that one important  condition for the asymmetric
photolysis by CPL to be at the origin of the terrestrial homochirality
of natural amino acids is generally overlooked, namely,  the
asymmetric photolysis should favour the L-enantiomer for {\it all} the
primordial amino acids involved in the genesis of life (i.e., biogenic
amino acids). Although this condition is probably satisfied for
aliphatic amino acids, some non-aliphatic amino acids like
tryptophan and proline may violate the condition and
thus invalidate the asymmetric photolysis scenario, assuming they were
among the primordial amino acids. Alternatively, if CPL photolysis in
space is indeed the source of homochirality of amino acids, then
tryptophan and proline may be crossed out from the list of biogenic
amino acids. Laboratory experiments suggested in this paper could shed
further light on the composition of the set of amino acids that were
required for development of the homochirality of first life. 
}

\bigskip   
\bigskip   
\bigskip   
 
\begin{center} 
{\normalsize\bf 1. The Origin of Amino Acid Homochirality:\\ 
A Long-Standing Question}
\end{center}
\medskip
Proteins play a crucial role in life, taking part
in all vital processes. The building blocks of proteins are the amino
acids. They consist of a central carbon atom (called
$\alpha$-carbon) bound to four groups: an amino or basic group
(NH$_2$), an acid group (COOH), a hydrogen atom, and a variable group
R, called side chain, that makes the specificity of each amino acid.
Only 20 different amino acids  are used as building blocks in today's
proteins; they constitute the set of so-called {\it natural} amino
acids. The primordial amino acids involved in the genesis of life
constitute the set of {\it biogenic} amino acids. In the following, we
will assume that some (if not all) of these primordial amino acids are
now part of the set of natural amino acids. 

Because the four chemical
groups bound to the $\alpha$-carbon of amino acids are not in a plane
but rather adopt a tetrahedral shape (Fig.~1), all amino acids are
chiral (except glycine whose R group is a hydrogen atom), i.e., they
possess two non-superposable three-dimensional mirror image structures
or enantiomers.  The refractive indices of (a solution of) chiral
molecules for clockwise and counterclockwise circularly polarized
light are different, leading to a net rotation of the plane of
linearly polarized light. By convention, molecules that make the
polarization plane of sodium D light (at $\lambda = 589.3$~nm) turn to
the right or to the left are called $(+)$ or $(-)$, respectively (note
that, for several amino acids, the $+/-$ assignment is different for
acid, neutral or basic solutions; see Sect.~3). The $+/-$ convention
thus classifies enantiomers on the ground of their {\it optical
rotatory power}. Several other classification schemes of enantiomers
exist, based on their {\it geometrical conformation}. One of these
compares the amino acid structure to that of a reference chiral
molecule, namely glyceraldehyde. By convention, $(+)$-glyceraldehyde
was assigned configuration D (from the Latin {\it dexter}, right), and
$(-)$-glyceraldehyde was assigned configuration L (from the Latin {\it
laevus}, left). Using some correspondence rule (see e.g., Morrison and
Boyd, 1987), the considered amino acid structure can be superposed on
either D- or L-glyceraldehyde, and it is classified as D or L
accordingly. Since the D/L classification refers to the geometrical
conformation whereas $+/-$ refers to the optical rotatory power, there
is not a one-to-one correspondence between the two assignments.
Actually, only a small majority of the 19 chiral natural L-amino acids
rotate the plane of polarized sodium light to the left, i.e., belong
to the L-($-$) type (in a neutral solution) (see e.g., Morrison and
Boyd, 1987). 

\begin{figure}[hbtp]
%\illu{98}{50}{Cerf_Jor.eps}
\resizebox{\hsize}{!}{{\includegraphics{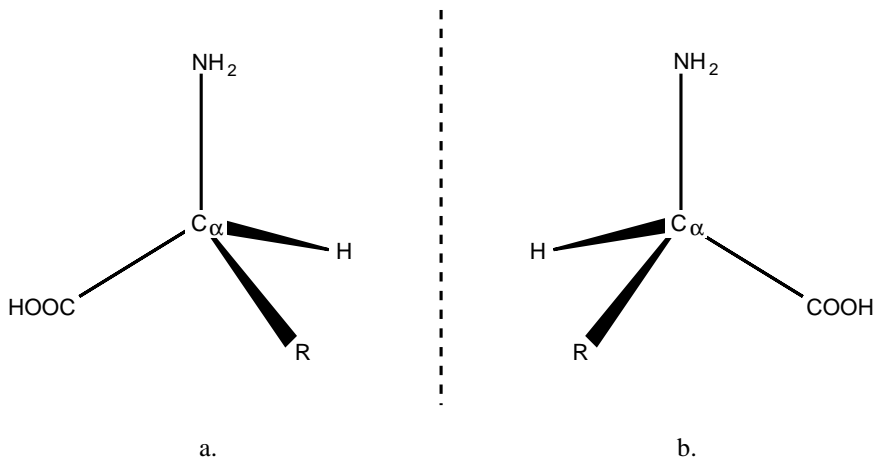}}}
\caption{a. L-amino acid; b. D-amino acid}
\label{fig:1}
\end{figure}

It is known since long ago that the natural amino acids are, with rare
exceptions, exclusively of the L-configuration (Davies, 1977). The
origin of this chiral homogeneity (i.e., {\it homochirality}) has been
a puzzle since its discovery, and remains the subject of a warm debate
(Cline, 1996; Podlech, 1999, and references therein). 
    
Several mechanisms have been proposed, as reviewed e.g. by
Bonner (1991) (see also the various contributions in Cline, 1996),
which may be grouped into biotic and abiotic theories. The former ones
assume that life originated on Earth through chemical evolution in a
primordial racemic (i.e., containing equal amounts of the L- and
D-enantiomers) milieu, and that chiral homogeneity inevitably resulted
from the evolution of living matter. Gol'danskii and Kuz'min (1988)
convincingly argued, however, that a biotic scenario for the origin of
chiral purity is not viable in principle, since without
preexisting chiral purity the self-replication characteristic of
living matter could not occur. This argument thus strongly
favours abiotic theories which may be grouped into the following
classes (Bonner, 1991): chance mechanisms (spontaneous symmetry
breaking by stereospecific autocatalysis, spontaneous resolution on
crystallization, asymmetric synthesis on chiral crystals,
asymmetric adsorption) and determinate mechanisms, the latter
being subdivided into regional/temporal processes
(symmetry breaking induced by electric, magnetic or gravitational
fields, by circularly polarized light via asymmetric
photoequilibration, photochemical asymmetric synthesis or
asymmetric photolysis) and universal processes (violation of
parity in the weak interaction).  

Only the hypothesis of symmetry breaking by the
action of circularly polarized light (CPL) 
will be discussed in some details in this paper. The role
of CPL present in natural skylight as a chiral engine was already
suggested in the nineteenth century by Le Bel (1874) and van
t'Hoff (1894). More recently, a similar scenario invoking the
asymmetric photolysis of amino acids taking place not on Earth but
rather in space (probably in the organic mantles at the surface of
interstellar grains) has been put forward
by several authors (Norden, 1977; Rubenstein et al.,
1983; Bonner and Rubenstein, 1987; Bonner, 1991; Bonner, 1992;
Greenberg et al., 1994; Greenberg, 1997). After some debate regarding
the possible role of CPL from pulsars (Rubenstein et al.,
1983; Roberts, 1984; Bonner and Rubenstein, 1987; Greenberg et al.,
1994; Engel and Macko, 1997; Mason, 1997; Bonner et al., 1999), this
idea regained interest recently  with the observation of CPL in the
Orion OMC-1 star-forming region (Bailey et al., 1998).

\bigskip   
 
\begin{center} 
{\normalsize\bf 2. Asymmetric Photolysis:\\ 
A Cosmic Enantioselective Engine?}
\end{center}
\medskip
Asymmetric photolysis, first
demonstrated successfully by Kuhn and coworkers (1929, 1930a, 1930b)
involves the preferential destruction of one enantiomer during the
photodegradation of a racemic mixture by CPL. CPL-mediated reactions
depend on the circular dichroism (CD) of the reactant (Crabb\'e, 1965;
Buchardt, 1974; Rau, 1983), i.e., on the difference in its molar
absorption coefficients $\epsilon_{(-)}$ and $\epsilon_{(+)}$ for
$(-)$-CPL (left CPL) and $(+)$-CPL (right CPL), respectively
[$\Delta\epsilon = \epsilon_{(-)} - \epsilon_{(+)} \ne 0$]. Note that
here the difference in {\it molar absorption coefficients} is
involved, not the difference in {\it refractive indices} (they are
nevertheless related via the Kramers-Kronig integral relation
(see e.g., Cantor and Schimmel, 1980). Since the photolysis rate
depends upon the amount of light absorbed by the reactant, CD thus
leads to different reaction rates for the two enantiomers, inducing an
enantiomeric asymmetry as the reaction proceeds. Asymmetric
photolysis, as considered here, results in the preferential
destruction of the enantiomer having the higher absorption
coefficient. The efficiency with which photolysis yields an
enantiomeric excess is directly related to the so-called anisotropy
factor $g = \Delta\epsilon/\epsilon$, where $\epsilon =
(\epsilon_{(-)} + \epsilon_{(+)})/2$ (Kuhn, 1930; Balavoine et al.,
1974). In the spectral region where $\Delta\epsilon \ne 0$, the
molecule is said to have a CD band, which corresponds to an absorption
band of the pertinent chromophore in the substrate.   
    
The electronic absorption bands of amino acids occur in the UV
(shortwards of 300~nm) (Donovan, 1969), and most are optically active. 
Successful asymmetric photolyses of amino acids were performed by
Flores et al. (1977) (2.5\% enantiomeric excess in leucine after 75\%
photolysis of a racemic mixture), Norden (1977) (0.22\% excess in
glutamic acid after 52\% photolysis, 0.06\% excess in
alanine after 20\% photolysis), and Greenberg et al. (1994) (3\%
excess after irradiating racemic tryptophan at 10~K for 50 hours with
monochromatic light at 252.4 nm with about $10^{12}$
photons~cm$^{-2}$~s$^{-1}$).
 
In his review on the {\it Origin and Amplification of
Biomolecular Chirality}, Bonner (1991) concludes that an
extraterrestrial origin of the biological homochirality on Earth
seems the most likely. This author suggests that an enantiomeric excess
in amino acids originated in space as a result of asymmetric
photolysis triggered by CPL (Norden, 1977; Rubenstein et al., 1983;
Bonner and Rubenstein, 1987; Greenberg et al., 1994; Greenberg, 1996), and
was somehow transported to the prebiotic Earth. This scenario received
further support
recently from the detection of IR CPL in the Orion OMC-1 star-
forming region (Bailey et al., 1998).

In order for such a scenario to work, several conditions must be met:
(i) amino acids must be able to form in an extraterrestrial
environment;
(ii) UV CPL must be present in space to irradiate these
extraterrestrial amino acids;   
(iii) all the biogenic L-amino
acids must have a CD spectrum such that the photolysis by CPL of a
given sign produces an excess of the L-enantiomer for all of
them; 
(iv) amino acids must be
transported from space onto the primitive Earth e.g. via cometary
and asteroidal impacts or via accretion of interstellar grains when
the Earth traverses molecular clouds. They must survive the heat
generated during the passage through the atmosphere and the impact 
with the surface;   
(v) amplification mechanisms are required to bring
the small excess induced by asymmetric photolysis to complete
homochirality.

Most of these conditions remain speculative, though with very
different levels of uncertainty. The proof of (i) would require
the direct detection of amino acids in space, which has not yet
been achieved with certainty, despite several
attempts (Miao et al., 1994; Travis, 1994, Combes et al., 1996).
Indirect arguments in favour of an exogenous synthesis of amino acids
are however provided by the discovery of apparently extraterrestrial
amino acids in the Cretaceous/Tertiary boundary sediments (Zhao and 
Bada, 1989; for critical assessments, see Cronin, 1989, Chyba et
al., 1990, Chyba and Sagan, 1992), and in the Murchison
meteorite (Kvenvolden et al., 1970; Engel and Nagy, 1982; Engel et al.,
1990, Engel and Macko, 1997; Cronin and Pizzarello, 1997, and
references therein; for a critical assessment, see Cronin and Chang,
1993, and Pizzarello and Cronin, 1998). Moreover, laboratory
experiments that simulate the formation of the organic mantles on
interstellar grains by the action of UV light have been able to
produce amino acids (Mendoza-G\'omez, 1992; Greenberg, 1997).

The detection of CPL in the Orion OMC-1 star-forming
region by Bailey et al. in 1998 (condition ii) gave a new
impetus to asymmetric photolysis in space as a cosmic enantioselective
engine. The large (17\%) circular polarization reported in Orion OMC-1
was observed in the IR domain, though amino acids CD bands are
located shortward of 300 nm. Nevertheless, model calculations
seem to indicate that, if CPL is produced by scattering on
nonspherical grains aligned in a magnetic field, similar circular
polarization levels should be attained in the UV and IR domains 
(Bailey et al, 1998). Somewhat lower circular polarization levels were
reported previously in the Chamaleon low-mass star-forming
region (Gledhill et al., 1996) and around the pre-main sequence object
GSS30 (Chrysostomou et al., 1997).

Transport of amino acids from space to the prebiotic Earth
(condition iv) and amplification of a small exogeneous
enantiomeric excess (condition v) seem possible as well, as shown
by detailed studies (Chyba et al., 1990; Bonner, 1991; Chyba and Sagan,
1992; Greenberg et al., 1994). 

Thus, only condition (iii), i.e., the possibility of forming the
same enantiomer for {\it all the biogenic amino acids} by
asymmetric photolysis, has not yet been the subject
of a critical assessment in order to validate
the above scenario (although a weaker form of this condition
was already expressed by Mason, 1997). In fact, the assessment of
condition (iii) requires the knowledge of both the composition of
the set of biogenic amino acids, and their CD spectrum in the
conditions prevailing in space (e.g., solid or gas phase,
temperature). Both of these are unknown, unfortunately. Nevertheless,
the consideration of CD spectra of natural amino acids in liquid
solution may already provide some useful information, as shown in
Sect.~3.

\bigskip   
 
\begin{center} 
{\normalsize\bf 3. CD Properties of Amino Acids}
\end{center}
\medskip
The possibility that the terrestrial homochirality of amino acids
originated from asymmetric photolysis by CPL in space requires that the
L-enantiomer be favoured for all the biogenic amino acids. In other
words, this requirement implies that the substrate was irradiated by
CPL in a spectral window where all the biogenic amino acids have a CD
band of {\it one and the same sign}.  As noted by Mason (1997) and
Bailey et al. (1998), an enantioselective effect on amino acids is
best obtained if the CPL spectrum is confined to a single CD band,
because CD bands alternate in sign and sum to zero over the whole
spectrum (the Kuhn-Condon rule: Kuhn, 1930; Condon, 1937). In the case
of broad-band CPL, a net enantioselective effect may nevertheless
result if the wavelength integral of the CD index weighted by the CPL
spectrum yields a non-zero effective CD coefficient
$\Delta\epsilon$ (Buchardt, 1974). To be at the origin of the
biomolecular homochirality, the (effective) CD coefficient must be of
the {\it same sign for all the biogenic amino acids}. 

CD data for amino acids may be found in Legrand and Viennet (1965,
1966), Myer and MacDonald (1967), Katzin and Gulyas (1968), Anand and
Hargreaves (1968), Horwitz et al. (1969), Sakota et al. (1970),
Fowden et al. (1971), as well as in the references quoted by 
Blout (1973). Their general properties, along with the
chromophore assignments, are summarized in Donovan (1969),
Crabb\'e (1971) and Blout (1973). As already mentioned, the CD data of
biogenic amino acids only should be examined in principle. However, as
the composition of the set of biogenic amino acids is not currently
known, we will discuss CD data of all natural amino acids, assuming
biogenic amino acids are among them. 

The optical activity of amino acids arises from the acid group
chromophore bound to their $\alpha$-carbon (i.e., a carboxyl group
COOH in acid medium, that deprotonates in a neutral or basic medium to
give a carboxylate group COO$^-$) and from possible supplementary
chromophores located in their side chain. The CD spectrum of aliphatic
amino acids (with side chains involving only C and H atoms without
double bonds, i.e., alanine, valine, leucine and isoleucine) is quite
simple, as it is due to the sole acid group chromophore bound to the
$\alpha$-carbon (Crabb\'e, 1971). By contrast, other amino acids
exhibit a more complex CD behaviour because they possess a
supplementary chromophore in their side chain (aromatic ring
for phenylalanine, tyrosine and tryptophan; sulfur-containing group
for cysteine and methionine; basic group for lysine, arginine and
histidine; acid group for aspartic and glutamic acids; side
chain closing back onto the $\alpha$-amino group for proline) 
(see Donovan, 1969; Blout, 1973).

Because the pH of the medium modifies the amino acid by protonating or
deprotonating the basic and acid groups (bound to the $\alpha$-carbon
or located in the side chain), the optical properties of amino
acids depend on the acidity of the medium and on the nature of the
solvent (Donovan, 1969). A sensitivity upon temperature (Horwitz et
al., 1969) and upon ionization state (Katzin and Gulyas, 1968) has also
been reported. Assuming that the acid group of amino acids in space
occur in the form of a carboxyl group COOH rather than of a 
carboxylate group COO$^-$ (since there is no reason for it to
deprotonate as in neutral or basic solutions), CD measurements in acid
medium should be considered when the optical properties of amino acids
are dominated by their carboxyl chromophore.

As indicated in the references quoted above, laboratory measurements
show that the carboxyl group bound to the $\alpha$-carbon has a strong
CD band centered at about 210 nm. The sign of this CD band is directly
related to the stereochemistry of the $\alpha$-carbon and is thus the
same for all L-amino acids. If this band were the only one involved in
the asymmetric photolysis, the photolysis of amino acids would indeed
favour the same enantiomer for all amino acids, and
extraterrestrial asymmetric photolysis could indeed be considered as a
viable explanation for the amino acid homochirality on Earth.

However, for non-aliphatic amino acids, the side chains complicate the
picture as they introduce supplementary chromophores. 
The situation appears especially critical with tryptophan, whose
indole chromophore exhibits a strong CD band centered at about
195~nm, with opposite sign to the carboxyl 210 nm
band (Legrand and Viennet, 1965; Myer and MacDonald, 1967; Blout, 1973). 
Proline also has a strong CD
band of opposite sign around 193 nm in
a neutral solution (this band however disappears in acid
solution; Fowden et al, 1971).

At this point, it should be stressed that Greenberg et al. (1994) have
obtained an enantiomeric excess starting from racemic tryptophan in
a laboratory experiment  simulating photolysis by CPL irradiating an
interstellar dust grain. The
experiment was conducted at a temperature of 10~K with monochromatic
light at 252.4~nm from a high pressure mercury lamp. 
Although that experiment certainly demonstrates the potential of CPL
to trigger  asymmetric photolysis of amino acids at the surface of
interstellar grains, it does not ensure that an irradiation
with broad-band UV CPL, as is more likely to be the case in space as
discussed by Bailey et al. (1998), would still result in an
enantiomeric excess (given the presence of CD bands of opposite
signs in tryptophan). Moreover, in order to ensure homochirality with
the other amino acids, the asymmetric photolysis of tryptophan should
be governed by the carboxyl chromophore (210~nm) rather than by the
indole chromophore (195~nm). This condition might be satisfied for an
irradiation by UV light from main sequence stars later than about A8,
but it does not necessarily hold true for irradiation by UV light
from earlier stars whose flux raises shortward of 200~nm (Bailey et al., 1998).
The same problem may arise in case of an irradiation by pulsar synchrotron
radiation with a constant $\lambda F_\lambda$ spectrum (Rubenstein et
al., 1983; Bonner and Rubenstein, 1987; Greenberg et al., 1994).

\bigskip   
 
\begin{center} 
{\normalsize\bf 4. Conclusion: Asymmetric Photolysis vs.
Composition\\of the Set of Biogenic Amino Acids} 
\end{center}
\medskip
The present paper has reviewed the conditions necessary for the
asymmetric photolysis of biogenic amino acids by CPL in space to be
at the origin of today's homochirality of natural amino acids. It has
shown that a critical requirement in that respect is that asymmetric
photolysis should select the L-enantiomer for {\it all} the
biogenic amino acids.

A survey of the available CD data for amino acids has revealed 
that tryptophan and proline pose the most serious problem, as they
exhibit CD bands of opposite signs in the UV region where early-type
main sequence stars emit most of their radiation.
Because the signs and intensities of CD bands depend however on the
properties of the medium, extrapolation of
laboratory data obtained in liquid solutions to infer the CD
properties of amino acids in space (where they are likely to be
found in solid or gas phase) is not straightforward, and prevents any
firm conclusion to be drawn at this stage.    
Asymmetric photolysis laboratory experiments along the guideline of
Greenberg et al. (1994), extended to other amino acids and
using broad band UV CPL rather than monochromatic light, would be of
great interest. As would be CD data for as many amino acids as
possible, obtained under experimental conditions
matching as closely as possible the conditions prevailing in space.

To summarize, the consideration of the available CD properties of amino
acids currently leads to two mutually exclusive possibilities regarding
the possible role of asymmetric photolysis in space for the
homochirality of the natural amino acids: (i) if tryptophan or
proline is biogenic, then the extraterrestrial
asymmetric photolysis scenario has to be rejected, or (ii) if that
scenario is valid, then the CD properties of amino acids (along
with the spectral properties of CPL in space) allow to eliminate
tryptophan and proline from the set of
biogenic amino acids. This reasoning could be extended to any other
amino acid for which new CD measurements in conditions mimicking those
in space would uncover CD bands with a sign opposite to that of the
carboxyl chromophore, in the spectral region characterizing CPL in
space.

It is currently impossible to decide between the two alternatives,
as there are experimental facts in support of each alternative. On
the one hand, proline has been found in the Murchison
meteorite (Kvenvolden et al., 1970; Engel and Nagy, 1982), which may
be indicative of its biogenic nature if life started from the amino
acids deposited on the early Earth, but on the other hand, some
authors (Isoyama et al., 1984) have argued that tryptophan may have
appeared quite late in the biological evolution. 

In conclusion,  we hope to have convinced the reader that
the role of extraterrestrial asymmetric photolysis in the origin of the
homochirality of natural amino acids on Earth, if at all involved, is
far more complicated than is usually apprehended in the
astronomical literature.

\newpage   
 
\begin{center} 
{\normalsize\bf Acknowledgments}
\end{center}
\medskip
We thank Prof. W. A. Bonner for sending us a paper in advance of publication.
A. J. is Research Associate of the FNRS
(Belgian National Fund for Scientific Research).

\bigskip

\begin{center}
{\normalsize\bf References}
\end{center}
\medskip
\begin{description}
{\footnotesize
\item Anand, R.D. and Hargreaves, M.K.: 1968,  {\it Chem. Ind.},
880. 
\item Bailey, J.,
Chrysostomou, A., Hough, J.H., Gledhill, T.M., McCall, A., Clark, S.,
M\'enard, F. and Tamura, M.: 1998,  {\it Science} {\bf 281}, 672.
\item Balavoine, G., Moradpour, A. and Kagan, H.B.: 1974,  
{\it J. Am. Chem. Soc.} {\bf 96}, 5152. 
\item Blout, E.R.: 1973,  in F. Ciardelli and
P. Salvadori (eds.), {\it Fundamental
Aspects and Recent Developments in ORD and CD}, Heyden \& Son,
London, p. 352. 
\item Bonner, W.A.: 1991,   {\it Origins of Life
Evol. Biosph.} {\bf 21}, 59.  
\item Bonner, W.A.: 1992,   {\it Origins
of Life Evol. Biosph.} {\bf 22}, 407. 
\item Bonner, W.A. and Rubenstein, E.: 1987,  {\it Biosystems}
{\bf 20}, 99.
\item Bonner, W.A., Rubenstein, E. and Brown, G.S.: 1999,  {\it 
Origins of Life Evol. Biosph.}{\bf 29}, 329-332. 
\item Buchardt, O.: 1974,  {\it Angew. Chem. Intern. Ed.} {\bf
13}, 179.   
\item Cantor, C.R. and Schimmel, P.R.: 1980, {\it 
Biophysical Chemistry, Part II}, Freeman and Co., San Francisco,
p.413. 
\item Chrysostomou, A., M\'enard, F., Gledhill, T.M.,
Clark, S., Hough, J.H., McCall,  A. and Tamura, M.: 1997,  {\it
MNRAS} {\bf 285}, 750.
\item Chyba, C.F. and Sagan, C.: 1992,   {\it Nature} {\bf 355},
125. \item Chyba, C.F., Thomas, P.J., Brookshaw, L. and Sagan, C.:
1990,   {\it Science} {\bf 249}, 366.
\item Cline, D.B. (ed.): 1996,  {\it Physical Origin of
Homochirality in Life} (AIP Conf. Proc. 379, American Institute of
Physics, Woodbury and New York).
\item Combes, F., Nguyen-Q-Rieu and Wlodarczak, G: 1996,  {\it
A\&A} {\bf 308}, 618.
\item Condon, E.U.: 1937,   {\it Rev. Mod. Phys.} {\bf 9}, 432.
\item Crabb\'e, P.: 1965,  in  {\it Optical Rotatory Dispersion and
Circular Dichroism in Organic Chemistry}, Holden-Day, San Francisco.
\item Crabb\'e, P.: 1971,  in  F.C. Nachod and J.J.
Zuckerman (eds.), {\it Determination of Organic
Structures by Physical Methods}, Academic Press, New York, p. 133.
\item Cronin, J.R.: 1989,  {\it Nature} {\bf 339}, 423.  
\item Cronin, J.R. and Chang, S.: 1993,  in J.M. Greenberg et al.
(eds.), {\it The Chemistry of Life's Origins}, Kluwer,
Dordrecht, p. 209. 
\item Cronin, J.R. and Pizzarello, S.: 1997,  {\it Science} {\bf
275}, 951.
\item Davies, J.S.: 1977,  in B. Weinstein (ed.), {\it
Chemistry and Biochemistry of Amino Acids, Peptides and Proteins},
Marcel Dekker, New York, p. 1. 
\item Donovan, J.W.: 1969,  in S.J. Leach (ed.),
{\it Physical Principles and Techniques of Protein Chemistry} (Part
A), Academic Press, New York, p.101. 
\item Engel, M.H. and Macko, S.A.: 1997,  {\it Nature} {\bf 389},
265. 
\item Engel, M.H., Macko, S.A. and Silfer, J.A.: 1990,  {\it
Nature} {\bf 348}, 47.
\item Engel, M.H. and Nagy, B.: 1982,   {\it Nature} {\bf 296},
837. \item Flores, J.J., Bonner, W.A. and Massey, G.A.: 1977, 
 {\it J. Am. Chem. Soc.} {\bf 99}, 3622. 
\item Fowden, L., Scopes, P.M. and Thomas, R.N.: 1971,  {\it J.
Chem. Soc. (C)}, 834.
\item Gledhill, T.M., Chrysostomou, A. and Hough, J.H.: 1996, 
{\it MNRAS} {\bf 282}, 1418.
\item Gol'danskii, V.I. and Kuz'min, V.V.: 1988,  {\it Z. 
Phys. Chem.} {\bf 269}, 216.
\item Greenberg, J.M.: 1996, 
in D.B. Cline (ed.) {\it Physical Origin of Homochirality in Life} (AIP
Conf. Proc. 379), American Institute of Physics, Woodbury and New
York, p.185.  
\item Greenberg, J.M.: 1997,  in  R.B. Hoover (ed.) {\it Instruments, 
Methods, and Missions for the Investigation of
         Extraterrestrial Microorganisms} (Proc. SPIE Vol. 3111), p. 226.
\item Greenberg, J.M., Kouchi, A., Niessen, W., Irth, H.,
Van Paradijs, J., de Groot, M. and Hermsen, W.: 1994,  {\it J.
Biol. Phys.} {\bf 20}, 61. 
\item Horwitz, J., Hardin Strickland, E. and
Billups, C.: 1969,  {\it J. Am. Chem. Soc.} {\bf 91}, 184.
\item Isoyama, M., Ohoka, H., Kikuchi, H., Shimada, A. and
Yuasa, S.: 1984,  {\it Origins of Life} {\bf 14}, 439. 
\item Katzin, L.I. and Gulyas, E.: 1968,  {\it J. Am. Chem.
Soc.} {\bf 90}, 247.
\item Kuhn, W.: 1930,  {\it Trans. Faraday Soc.} {\bf 26}, 293.
\item Kuhn, W. and Braun, E.: 1929,  {\it Naturwissenschaften}
{\bf 17}, 227.
\item Kuhn, W. and Knopf, E.: 1930a,  {\it Z. Physik. Chem.}
{\bf B 7}, 292. 
\item Kuhn, W. and Knopf, E.: 1930b, 
{\it Naturwissenschaften} {\bf 18}, 183.
\item Kvenvolden, K.A., Lawless, J., Pering, K., Peterson, E.,
Flores, J., Ponnamperuma, C., Kaplan, I.R. and Moore, C.: 1970,  
{\it Nature} {\bf 228}, 923.
\item Le Bel, J.A.: 1874,  {\it Bull. Soc. Chim. France} {\bf
22}, 337.  
\item Legrand, M. and Viennet, R.: 1965,  {\it Bull. Soc.
Chim. France}, 679. 
\item Legrand, M. and Viennet, R.: 1966, 
{\it Bull. Soc. Chim. France}, 2798.
\item Mason, S.F.: 1997,  {\it Nature} {\bf 389}, 804.
\item Mendoza-G\'omez, C.X.: 1992, {\it Complex Irradiation Products
in the Interstellar Medium}, Ph.D. thesis (University of Leiden).  
\item Miao, Y., Snyder, L.E., Kuan, Y.J. and
Lovas, F.J.: 1994,  {\it BAAS} {\bf 26}, 906.
\item Morrison, R.T. and Boyd, R.N.: 1987, {\it Organic Chemistry},
Allyn and Bacon, Boston.
\item Myer, Y.P. and MacDonald, L.H.: 1967,  {\it J. Am. Chem.
Soc.} {\bf 89}, 7142.
\item Norden, B.: 1977,  {\it Nature} {\bf 266}, 567.
\item Pizzarello, S. and Cronin, J.R.: 1998,  {\it Nature} {\bf
394}, 236. 
\item Podlech, J.: 1999,  {\it Angew. Chem. Int. Ed.} {\bf
38}, 477. 
\item Rau, H.: 1983,  {\it Chem. Rev.} {\bf 83}, 535.
\item Roberts, J.A.: 1984,  {\it Nature} {\bf 308}, 318.
\item Rubenstein, E., Bonner, W.A., Noyes, H.P. and Brown, G.S.: 1983, 
{\it Nature} {\bf 306}, 118.
\item Sakota, N., Okita, K. and Matsui, Y.: 1970,  {\it Bull. 
Chem. Soc. Japan} {\bf 43}, 1138.
\item Travis, J.: 1994,  {\it Science} {\bf 264}, 1668.
\item van't Hoff, J.H.: 1894,  in {\it The Arrangement of Atoms in
Space}, 2nd ed., Braunschweig, p. 30. 
\item Zhao, M. and Bada, J.L.: 1989,  {\it Nature} {\bf 339},
463.

} 
\end{description}

\end{document}